\documentclass[aps,prc,twocolumn,superscriptaddress,showpacs,floatfix,nofootinbib]{revtex4-1}
\usepackage{amsmath,graphicx}
\newcommand {\av}[1]       {\ensuremath{\left \langle #1 \right \rangle}}
\newcommand {\rsq}         {\mbox{$\langle r^2 \rangle$}}
\newcommand {\es}          {\ensuremath{\varepsilon_{s}}}
\begin{document}

\title{Understanding anisotropy generated by fluctuations 
  in heavy-ion collisions}

\author{Rajeev S. Bhalerao}
\affiliation{Department of Theoretical Physics, TIFR,
   Homi Bhabha Road, Colaba, Mumbai 400 005, India}
\author{Matthew Luzum}
\affiliation{
CEA, IPhT, Institut de physique theorique de Saclay, F-91191
Gif-sur-Yvette, France} 
\author{Jean-Yves Ollitrault}
\affiliation{
CNRS, URA2306, IPhT, Institut de physique theorique de Saclay, F-91191
Gif-sur-Yvette, France} 
\date{\today}

\begin{abstract}
Event-by-event fluctuations are central to the current understanding
of ultrarelativistic heavy-ion collisions. 
In particular, fluctuations in the geometry of the early-time collision system
are responsible for new phenomena such as triangular flow, which have solved important puzzles in existing data. 
We propose a simple model 
where initial fluctuations stem from independent flux tubes randomly 
distributed in the transverse plane.  
We calculate analytically the moments of the
initial anisotropies (dipole asymmetry, eccentricity, triangularity),
which are the sources of anisotropic flow, and their mutual
correlations. 
Our analytic results are in good agreement with calculations from
commonly-used Monte-Carlo codes, providing a simple understanding of the fluctuations contained in these models.  Any deviation from these results in future experimental data would thus indicate the presence of non-trivial correlations between the initial flux tubes and/or extra sources of fluctuations that are not present in current models.
\end{abstract}


\maketitle
\section{Introduction}

Fluctuations of the positions of nucleons within nuclei result
in event-by-event fluctuations of the density of matter produced in
ultrarelativistic nucleus-nucleus
collisions~\cite{Aguiar:2001ac,Mrowczynski:2002bw,Miller:2003kd}.  
These event-by-event fluctuations have been observed through the
centrality and system-size dependence of elliptic 
flow~\cite{Alver:2006wh}, which is the second Fourier harmonic of the
azimuthal distribution of outgoing
particles. 
They may generate elliptic flow even in high-multiplicity
proton-proton collisions at the 
LHC~\cite{CasalderreySolana:2009uk,Avsar:2010rf}, 
which might be responsible~\cite{Bozek:2010pb,Werner:2010ss}
for the near-side ridge observed by CMS~\cite{Khachatryan:2010gv}. 
Furthermore, fluctuations result in new, odd harmonics 
of the azimuthal distribution of outgoing particles, which are clearly
seen experimentally through two-particle
correlations~\cite{Luzum:2010sp,Luzum:2011mm}:
The third harmonic, triangular flow~\cite{Alver:2010gr}, is
largely responsible for the  ridge and away-side structure seen in
two-particle correlations and has recently been analyzed at
RHIC~\cite{Adare:2011tg} and LHC~\cite{Aamodt:2011vk}. 
A smaller first harmonic, 
directed flow~\cite{Teaney:2010vd,Luzum:2010fb}, is also predicted. 
Finally, there are non-trivial correlations between these three
harmonics~\cite{Staig:2010pn,Teaney:2010vd}, and they can be directly
measured experimentally~\cite{Bhalerao:2011yg}.

Flow fluctuations are typically studied within Monte-Carlo models of
nucleus-nucleus collisions including initial-state fluctuations, such
as NeXSPheRIO~\cite{Andrade:2006yh}, UrQMD coupled to
hydrodynamics~\cite{Petersen:2010cw}, EPOS coupled to hydrodynamics~\cite{Werner:2011fd}, AMPT~\cite{Xu:2010du,Xu:2011fe},
and Glauber models coupled to
hydrodynamics~\cite{Broniowski:2007ft,Schenke:2010rr}.  
These complex Monte-Carlo approaches do not allow for a simple
understanding of the key quantities driving flow fluctuations. 

In this paper, we propose a simple model of initial-state fluctuations
and, by comparing with full Monte-Carlo calculations, show that it 
captures the essential physics contained in common models. We assume that the initial 
energy density profile in the transverse plane is the superposition of
$N$ random independent~\cite{Bhalerao:2006tp} and identical sources, namely: 
\begin{equation}
\label{themodel}
\epsilon(\vec x)=\sum_{j=1}^N \rho\left(\left|\vec x-\vec x_j\right|\right),
\end{equation}
where $\vec x_j$ are $N$ independent random variables with a 
smooth probability distribution $p(\vec x_j)$, and $\rho(r)$ is the
profile of a single source. 
The coordinate system is chosen such that centers of colliding nuclei
are on the $x$ axis, and the origin lies halfway between the two
centers.
For a collision of identical nuclei, 
the system then has the symmetries $p(x,-y)=p(x,y)$ and  $p(-x,-y)=p(x,y)$.  
A standard choice for $\rho(r)$ is a Gaussian
profile~\cite{Holopainen:2010gz,Qin:2010pf,Broniowski:2007ft,Schenke:2010rr} 
\begin{equation}
\label{gaussiansource}
\rho(r)=\rho_0 e^{-r^2/\sigma^2},
\end{equation}
where $\sigma$ controls the width. 
The parameters in our model are $N$, the number of sources, the
source distribution $p(\vec x)$ and the source profile $\rho(r)$. 

In Sec.~\ref{s:cumulantsandmoments}, we introduce the initial
anisotropies $\varepsilon_n$ and their moments, which are the relevant 
quantities. We discuss to what extent $\varepsilon_n$ depends on the
source profile $\rho(r)$.
In Sec.~\ref{s:2ndorder}, we derive analytic expressions for the rms
$\varepsilon_n$ and compare our results with two Monte-Carlo models.
In Sec.~\ref{s:higherorder}, our results are extended to higher
moments of the distribution of $\varepsilon_n$ and mixed correlations
between harmonics~\cite{Staig:2010pn,Teaney:2010vd,Bhalerao:2011yg}.

\section{Quantifying initial fluctuations}
\label{s:cumulantsandmoments}

\subsection{Azimuthal asymmetries of the initial distribution}
\label{s:cumulants}
In a given event, the azimuthal distribution of outgoing
particles is driven by the azimuthal distribution of the initial 
density~\cite{Holopainen:2010gz,Gardim:2011qn,Qin:2010pf}. 
We define the participant eccentricity~\cite{Alver:2006wh}
$\varepsilon_2$, the triangularity $\varepsilon_3$, 
and the dipole asymmetry~\cite{Teaney:2010vd} $\varepsilon_1$, and the corresponding
orientations $\Phi_n$, by:
\begin{eqnarray}
\label{defepsn}
\varepsilon_2 e^{2i\Phi_2} &\equiv & - \frac{\{r^2e^{2i\phi}\} }{\{r^2\}}\cr
\varepsilon_3 e^{3i\Phi_3} &\equiv & - \frac{\{r^3e^{3i\phi}\} }{\{r^3\}}\cr 
\varepsilon_1 e^{i\Phi_1} &\equiv & - \frac{\{r^3e^{i\phi}\} }{\{r^3\}},
\end{eqnarray}
where $\{\cdots\}$ denotes an average over the transverse plane in a
single event, weighted with the energy density:
\begin{equation}
\{f(x,y)\}\equiv
\frac{\int{f(x,y)\epsilon(x,y)dxdy}}{\int{\epsilon(x,y)dxdy}},
\end{equation}
and $(r,\phi)$ are polar
coordinates in which the pole is the center of the distribution, that
is, $\{r e^{i\phi}\}=0$. 

The numerators in the right-hand side of Eq.~(\ref{defepsn}) are the
cumulants introduced by Teaney and Yan~\cite{Teaney:2010vd}. 
The anisotropic flow of outgoing particles is driven by these initial
anisotropies, in the sense that the relation
$v_n\propto\varepsilon_n$ holds approximately~\cite{Alver:2010gr,Holopainen:2010gz,Qin:2010pf,Gardim:2011qn,Qiu:2011iv}. 

\subsection{Dependence of $\varepsilon_n$ on source size}
\label{s:size}

Within our simple model, one can easily study the dependence of
$\varepsilon_n$ on the profile of a 
single source $\rho(r)$. 
This is relevant to the early-time dynamics, which has a smearing
effect on each source~\cite{Qin:2010pf} and widens the source
profile. 

The generating function of cumulants is $W(\vec k)$, where~\cite{Teaney:2010vd} 
\begin{equation}
e^{W(\vec k)}\equiv \left\{ e^{i\vec k\cdot\vec x}\right\}.
\end{equation}
Teaney and Yan's cumulants are obtained by expressing $W(k_x,k_y)$ in
terms of $K\equiv k_x+ik_y$ and $\bar K\equiv k_x-ik_y$, and expanding
in power series of $K^p\bar K^q$. 
Specifically, the numerators of $\varepsilon_{1}$, $\varepsilon_{2}$
and  $\varepsilon_{3}$ in Eq.~(\ref{defepsn}) are obtained by
expanding $W(K,\bar K)$ to order $K\bar K^2$, $\bar K^2$, and $\bar K^3$,
respectively, and $\{r^2\}$ is obtained by expanding $W(K,\bar K)$ to order 
$K\bar K$.

In our model, Eq.~(\ref{themodel}), the density profile, is the
convolution of the profile of a single source with 
the distribution of sources $\sum_j\delta(\vec x-\vec x_j)$. Hence
\begin{equation}
\label{convolution}
W(\vec k)=\ln\left(\frac{\tilde\rho(\vec k)}{\tilde\rho(\vec 0)}\right)
+\ln\left(\frac{1}{N}\sum_{j=1}^N e^{i\vec 
  k\cdot\vec x_j}\right),
\end{equation}
where $\tilde\rho(\vec k)$ is the Fourier transform of the source
profile $\rho(r)$. By symmetry, $\tilde\rho(\vec k)$  depends only on
$|\vec k|^2=K\bar K$. Therefore the numerators of $\varepsilon_n$ in
Eq.~(\ref{defepsn}) are
strictly insensitive to the profile $\rho(r)$. The only dependence of
$\varepsilon_n$ on the source profile is contained in the denominator,
$\{r^n\}$.
If the source has a finite size, $\{r^n\}$ increases, which results in
a smearing of anisotropies. 
One expects that $\{r^n\}$ scales approximately like $\{r^2\}^{n/2}$, 
resulting in a stronger smearing  for $\varepsilon_3$ than for $\varepsilon_2$,
as observed in numerical calculations~\cite{Qin:2010pf}. 
One also expects that 
$\varepsilon_1/\varepsilon_3$ and $\varepsilon_3/\varepsilon_2^{3/2}$
are largely independent of the source profile.

\subsection{Moments}
\label{s:moments}
In practice, anisotropic flow is not analyzed in a single event, but
in a centrality class. It is inferred from multiparticle azimuthal
correlations~\cite{Borghini:2001vi}. To the extent that $v_n$ is
driven by $\varepsilon_n$, the measured azimuthal correlations scale
like the corresponding moments of the distribution of
$\varepsilon_n$ and their joint correlations. We introduce the
notation~\cite{Bhalerao:2011yg}:
\begin{eqnarray}
\label{moments}
\varepsilon\{n_1,\ldots,n_k\}
&\equiv& 
\left\langle\varepsilon_{n_1}e^{in_1\Phi_{n_1}}\ldots\varepsilon_{n_k}e^{i n_k\Phi_{n_k}}
\right\rangle,
\end{eqnarray}
where 
angular brackets denote an average over events in a centrality class. 
With this notation, the rms average of $\varepsilon_n$ is 
\begin{equation}
\varepsilon_n\{2\}\equiv \langle\varepsilon_n^2\rangle^{1/2}=
\varepsilon\{n,-n\}^{1/2}, 
\end{equation}
where we have used the notation $\varepsilon_n\{2\}$ for the rms value
of $\varepsilon_n$~\cite{Miller:2003kd}.

Symmetry with respect to the reaction plane ($y\to -y$) implies that
all moments are real, therefore 
$\varepsilon\{-n_1,\cdots,-n_k\}=\varepsilon\{n_1,\cdots,n_k\}$. 
In addition, 
$(x,y)\to (-x,-y)$ symmetry implies that the only nonvanishing moments are
those such that $n_1+\cdots+n_k$ is even. 
Since the azimuthal orientation of each collision is uncontrolled
experimentally, one can only measure rotationally symmetric
quantities. Thus, the only relevant moments are those satisfying
$n_1+\cdots+n_k=0$. 
(Some of our intermediate results will involve other
moments, such as $\varepsilon\{1,1\}$ or $\varepsilon\{3,-1\}$.)

Within our independent-source model, all moments can be calculated
analytically as a systematic expansion in powers of $1/N$, where $N$ is the number of sources.. 

\section{Rms asymmetries}
\label{s:2ndorder}

The rms value of $\varepsilon_n$ is of direct relevance to analyses of
anisotropic flow. Indeed, the simplest measurement of anisotropic flow
is a pair 
correlation, $\langle\cos n\Delta\phi\rangle$, which is the average
value over events of $v_n^2$. Assuming that $v_n\propto\varepsilon_n$ on an
event-by-event basis, this scales like 
$\langle\varepsilon_n^2\rangle=\varepsilon_n\{2\}^2$. 
In this Section, we derive analytic results for 
$\varepsilon_n\{2\}$ within our independent-source model, and
compare with numerical results. 

\subsection{Analytic results}

We denote by $\langle f(x,y)\rangle$ the average value of $f(x,y)$
with the source probability density $p(\vec x)$, and 
we introduce the notation $\delta_f\equiv \{f\}-\langle f\rangle$
for the event-by-event fluctuations.
We use the complex coordinate $z=x+iy$. The asymmetry 
$\varepsilon_n$ is given by Eq.~(\ref{defepsn}), where we replace
$re^{i\phi}$ by $z-\delta_z$ to take into account the recentering
correction. To leading order in fluctuations, one obtains
\begin{eqnarray}
\label{loepsn}
\varepsilon_3 e^{3i\Phi_3} &=& - \frac{\{(z-\delta_z)^3\} }{\{r^3\}}
\simeq - \frac{\delta_{z^3}-3\langle z^2\rangle\delta_z }{\langle
  r^3\rangle }\cr 
\varepsilon_1 e^{i\Phi_1} &=& - \frac{\{(z-\delta_z)^2(\bar
  z-\delta_{\bar z})\} }{\{r^3\}}\cr
&\simeq& - \frac{\delta_{z^2\bar z}-2\langle z\bar
  z\rangle\delta_z-\langle z^2\rangle\delta_{\bar z}}{\langle
  r^3\rangle },
\end{eqnarray}
where $\bar z=x-iy$. 
The rms value of $\varepsilon_n$ involves an average over events of
products of $\delta$'s. 
Two-point averages are computed using the following identity, which
holds for independent sources~\cite{Bhalerao:2006tp}:
\begin{equation}
\label{2point}
\langle\delta_f\delta_g\rangle=\frac{\langle fg\rangle-\langle
  f\rangle\langle g\rangle}{N}.
\end{equation}
We thus obtain, using the identities $\langle z^n\bar z^m\rangle=0$ 
for odd $n-m$ and  $\langle z^n\bar z^m\rangle=\langle
r^{n+m}\cos((n-m)\phi)\rangle$ for even $n-m$:
\begin{eqnarray}
\varepsilon_3\{2\}^2
&=& \frac{\langle r^6\rangle + 6\varepsilon_s \langle r^2\rangle \langle r^4 
\cos 2\phi\rangle + 9 \varepsilon^2_s \langle r^2\rangle^3}
{N\langle r^3\rangle^2} \cr
\varepsilon_1\{2\}^2
&=&\frac{1}{
N \langle r^3\rangle^2} 
\left[\langle r^6\rangle - 4\langle r^2\rangle 
\langle r^4\rangle + 4\langle r^2\rangle^3 \right. \cr
&& \left. + 2\varepsilon_s \langle r^2\rangle \langle r^4 \cos 2\phi\rangle + 
5 \varepsilon_s^2 \langle r^2\rangle^3\right].
\label{rmsepsilon}
\end{eqnarray}
In these equations, 
$\varepsilon_s\equiv -\langle r^2\cos 2\phi\rangle/\langle r^2\rangle$
denotes the standard eccentricity. 
We also recall the result for $\varepsilon_2\{2\}$ which has been 
derived earlier~\cite{Bhalerao:2006tp}. 
\begin{equation}
\label{eps2}
\varepsilon_2\{2\}^2
=\varepsilon_s^2+\frac{\langle
  r^4\rangle(1+3\varepsilon_s^2)+4\varepsilon_s\langle r^4\cos
  2\phi\rangle}{N\langle r^2\rangle^2}.
\end{equation}
The first term in the right-hand side is the standard eccentricity,
and the second term is the contribution of eccentricity fluctuations
to order $1/N$. 
For a large number of sources, $N\gg 1$, 
the participant eccentricity $\varepsilon_2$ 
reduces to the standard eccentricity, while odd harmonics
$\varepsilon_1$ and $\varepsilon_3$ vanish.

\subsection{Comparison with Monte-Carlo results}
\label{s:results}

We now compare analytic results derived from our
independent-source model with results obtained using the mckt-v1.00 Monte-Carlo~\cite{mckt}.
With this Monte-Carlo one can calculate results from both a Color-Glass-Condensate (CGC) inspired model --- the MC-KLN~\cite{Drescher:2007ax} improved with 
running-coupling BK unintegrated gluon densities~\cite{Albacete:2010ad}, as well as a standard
Monte-Carlo Glauber~\cite{Miller:2007ri}.
We present only results for Pb-Pb 
collisions at 2.76~TeV per nucleon-nucleon collision, though results
for 200~GeV Au-Au collisions agree equally well. 
In the Glauber model, each participant nucleon is given a
weight~\cite{Gombeaud:2009am} 
$w=1-x+xN_{\rm coll}$, where $N_{coll}$ is the number of binary
collisions of that nucleon, and $x=0.18$~\cite{Bozek:2010wt}. 

One input of our model is the probability distribution of sources in
the transverse plane, $p(\vec x)$. For the sake of consistency, we assume
that sources are distributed according to the average density profile: 
$p(\vec x)\equiv \langle\epsilon(\vec x)\rangle$, where
$\langle\cdots\rangle$ denotes an average over many events in a
centrality class. 
We assume pointlike sources for simplicity: $\rho(|\vec
x|)=\delta(\vec x)$. 
The last free parameter in our model is the number of independent
sources $N$.
One expects that this number scales typically like the number of
participant nucleons in a collision. However, participants are not
independent, but strongly correlated: for each participant of the
projectile, there is by definition at least one participant from the
target which is close enough in the transverse plane for a collision
to occur.  Nevertheless, it is plausible that the system behaves like a set of $N$ independent clusters of two or more nucleons.  Again for simplicity, we use $N=0.45N_{\rm part}$ for all
centralities, though one could make the agreement with Monte-Carlo even better
by fitting $N$ for each centrality class. 
Increasing $N$ by 20\%  typically decreases $\varepsilon_3\{2\}$ and
$\varepsilon_1\{2\}$ by 10\%, and $\varepsilon_2\{2\}$ by less than 4\%. 
If one uses a larger value of $N$ for peripheral collisions (say,
$N=0.6N_{\rm part}$ instead of $N=0.45N_{\rm part}$), agreement  with Monte-Carlo is significantly better for 
$\varepsilon_1\{2\}$ and $\varepsilon_3\{2\}$ but slightly worse for
$\varepsilon_2\{2\}$.

\begin{figure}[ht]
\includegraphics[width=\linewidth]{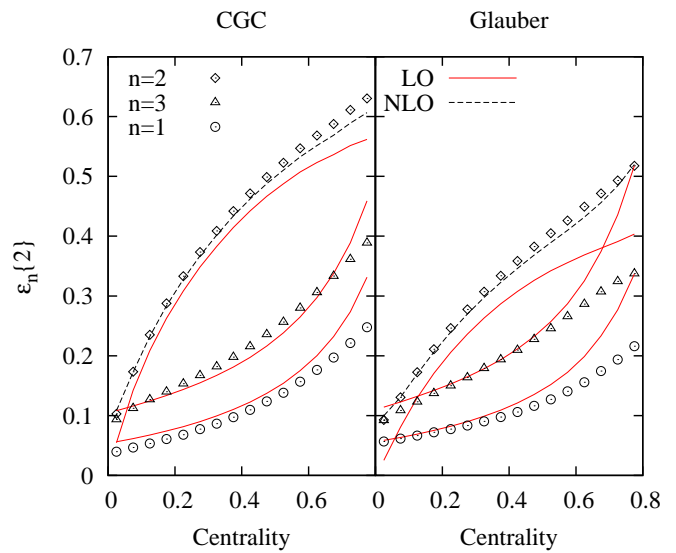}
\caption{(Color online) 
$\varepsilon_n\{2\}$, with $n=1,2,3$, versus centrality.  Symbols are Monte-Carlo results, lines are our
analytic results for independent 
sources. 
For $\varepsilon_2\{2\}$, the dashed line is the full result to order
$1/N$ (Eq.~(\ref{eps2})), while the solid line is the standard
eccentricity (first term in the right-hand side of Eq.~(\ref{eps2})).
}
\label{fig:epsn2}
\end{figure}
Fig.~\ref{fig:epsn2} displays a comparison between our 
result (\ref{rmsepsilon}) and Monte-Carlo calculations. 
Our analytic formulae reproduce the centrality dependence of
$\varepsilon_n$ computed using Monte-Carlo KLN or Monte-Carlo
Glauber\footnote{We have also done comparisons with 
  the PHOBOS  Monte-Carlo Glauber~\cite{Alver:2008aq}, and found that
  $\varepsilon_1$ from this model is significantly smaller than 
  predicted by our analytic formula.}.
Odd harmonics, $\varepsilon_1$ and $\varepsilon_3$, have a mild
centrality dependence and essentially scale like $1/\sqrt{N_{\rm part}}$. 
On the other hand, $\varepsilon_2$ is much larger for semi-central
collisions: this increase is driven by the almond shape of the
overlap area between the two nuclei. 
For $\varepsilon_2$, we show both the standard eccentricity, which is
the leading-order term in a $1/N$ expansion, and the full expression
including eccentricity fluctuations to order $1/N$ (Eq.~(\ref{eps2})). 
The standard eccentricity is significantly larger for KLN than for
Glauber~\cite{Hirano:2005xf,Lappi:2006xc}, so that eccentricity
fluctuations (which are roughly the same in both models) are a smaller
relative correction. 

Our model explains why $\varepsilon_1<\varepsilon_3$, a
feature which was observed in Monte-Carlo calculations but yet
unexplained. 
This may be readily understood from Eq.~(\ref{rmsepsilon}) for central 
collisions, where $\varepsilon_s=0$: only the $\langle r^6\rangle$
remains for $\varepsilon_3\{2\}^2$, while $\varepsilon_1\{2\}^2$ has an
additional contribution proportional to $\langle r^2\rangle^2-\langle
r^4\rangle$, which is negative. 
For sake of completeness, we have also carried out two other sets 
of Glauber calculations with only wounded nucleon scaling ($x=0$) or
binary collision scaling ($x=1$). Values of
$\varepsilon_1/\varepsilon_3$ turn out to be significantly smaller for
wounded nucleons (between 0.3 and 0.4 for 0-50\% centralities) than
for binary collisions (between 0.6 and 0.7). This strong ordering
is not reproduced by our analytic result, which gives intermediate 
results (between 0.5 and 0.6 in both cases).  
We do not have a simple explanation for this discrepancy.

\section{Higher-order moments}
\label{s:higherorder}

\subsection{Fluctuations of $\varepsilon_n$} 

The ALICE collaboration has recently measured
$v_2\{4\}$~\cite{Aamodt:2010pa}
 and 
$v_3\{4\}$~\cite{Aamodt:2011vk}. 
The relative magnitude of $v_n\{4\}$ and $v_n\{2\}$ depends on 
event-by-event fluctuations of $v_n$ if nonflow effects are 
small. Assuming that $v_n$ is proportional to $\varepsilon_n$ on an
event-by-event basis, fluctuations of $v_n$ are due to fluctuations of
$\varepsilon_n$:
\begin{equation}
\label{4cumulant}
\left(\frac{v_n\{4\}}{v_n\{2\}}\right)^4=
\left(\frac{\varepsilon_n\{4\}}{\varepsilon_n\{2\}}\right)^4\equiv 
2-\frac{\langle\varepsilon_n^4\rangle}{\langle\varepsilon_n^2\rangle^2}, 
\end{equation}
where we have introduced the 4-cumulant $\varepsilon_n\{4\}$ 
\cite{Miller:2003kd}. 
$v_n\{4\}$ is thus related to
$\langle\varepsilon_n^4\rangle$, which is $\varepsilon\{n,n,-n,-n\}$
in the notation of Eq.~(\ref{moments}). 
\begin{figure}[ht]
\includegraphics[width=\linewidth]{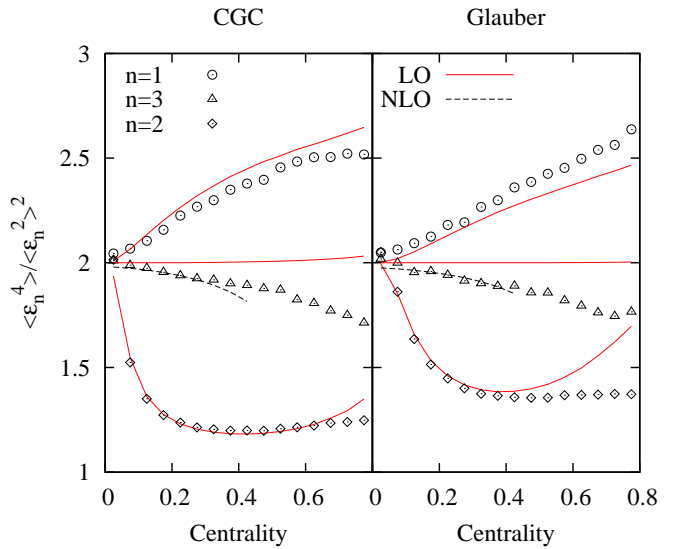}
\caption{(Color online) 
Ratio $\langle\varepsilon_n^4\rangle/\langle\varepsilon_n^2\rangle^2$, 
with $n=1,2,3$, versus centrality. 
As in Fig.~\ref{fig:epsn2}, symbols are Monte-Carlo results. 
Full lines are leading-order analytic results. 
The dashed line for $n=3$ includes next-to-leading order corrections
derived in Appendix. Although our next-to-leading order result is
valid for central collisions only, we compare with Monte-Carlo data
up to 40\% centrality. 
}
\label{fig:epsn4}
\end{figure}
Fig.~\ref{fig:epsn4} displays the  ratio
$\langle\varepsilon_n^4\rangle/\langle\varepsilon_n^2\rangle^2$
for 
$n=1,2,3$. The larger the ratio, the larger the fluctuations of
$\varepsilon_n$. 
Note that the order of the curves in Fig.~\ref{fig:epsn4} 
is reversed compared to Fig.~\ref{fig:epsn2}:
smaller values of $\varepsilon_n$ go with larger fluctuations. 
For the most central collisions, ratios are close to $2$ for all
$n$. This value corresponds to Gaussian
fluctuations~\cite{Voloshin:2007pc}.

For $n=2$, the ratio quickly decreases with centrality. This is due to
the standard eccentricity $\varepsilon_s$, which is zero for central
collisions but dominates over eccentricity fluctuations above 10\% 
centrality. 
The smaller eccentricity fluctuations, the closer the ratio 
to $1$. 
The decrease is stronger for the KLN model than for the Glauber
model because of the larger $\varepsilon_s$. 
Monte-Carlo results are compared with an analytic result 
using the expression of $\varepsilon_2\{4\}$ to order $1/N$ derived
in~\cite{Bhalerao:2006tp}.  
In practice, $\varepsilon_2\{4\}\simeq\varepsilon_s$ for all 
centralities. 

We now compute $\langle\varepsilon_3^4\rangle$ within our
independent-source model. 
To leading order in the fluctuations, $\varepsilon_3$ is given by 
Eq.~(\ref{loepsn}), and  $\langle\varepsilon_3^4\rangle$ involves 
products of four $\delta$'s. To leading order in $1/N$, average values
of such products can be expressed in terms of two-point averages
(Eq.~(\ref{2point}))  using Wick's theorem. For instance, the
four-point function is~\cite{Alver:2008zza} 
\begin{equation}
\langle\delta_{f}\delta_{g}\delta_{h}\delta_{k}\rangle=
\langle\delta_{f}\delta_{g}\rangle\langle\delta_{h}\delta_{k}\rangle+
\langle\delta_{f}\delta_{h}\rangle\langle\delta_{g}\delta_{k}\rangle+
\langle\delta_{f}\delta_{k}\rangle\langle\delta_{g}\delta_{h}\rangle. 
\end{equation}
This identity gives 
\begin{equation}
\varepsilon\{3,3,-3,-3\}=2\varepsilon\{3,-3\}^2+\varepsilon\{3,3\}\varepsilon\{-3,-3\}, 
\end{equation}
or, equivalently, 
\begin{equation}
\label{eps34}
\frac{\langle\varepsilon_3^4\rangle}{\langle\varepsilon_3^2\rangle^2}
=2+\left(\frac{\varepsilon\{3,3\}} {\varepsilon_3\{2\}^2}\right)^2\ .
\end{equation}
$\varepsilon\{3,3\}$ can easily be calculated to order $1/N$
in the same way as $\langle\varepsilon_3^2\rangle$:
\begin{eqnarray}
\label{eps33}
\varepsilon\{3,3\}&=&\langle\varepsilon_3^2\cos 6\Phi_3\rangle\cr
&=& \frac{1}{N\langle r^3\rangle^2}\left(
\langle r^6\cos 6\phi\rangle
+6\varepsilon_s \langle r^2\rangle \langle r^4 \cos 4\phi\rangle 
\right. \cr &&\left. 
- 9 \varepsilon_s^3 \langle r^2\rangle^3\right).
\end{eqnarray}
Inserting Eqs.~(\ref{rmsepsilon}) and (\ref{eps33}) into 
(\ref{eps34}), we obtain a leading-order analytic result for the
ratio. 
This analytic result is independent of $N$, and very close to 2 in
practice, as can be seen in Fig.~\ref{fig:epsn4}. 
This means that the first term on the right-hand side is the dominant 
contribution~\cite{Voloshin:2007pc}. 
It is easy to understand why the second term is small: 
$\varepsilon\{3,3\}$ involves the $6^{\rm th}$ Fourier harmonic of the
initial distribution, which is of order $\varepsilon_s^3$. Therefore
the last term in Eq.~(\ref{eps34}) is of order $\varepsilon_s^6$,
which is very small in practice. 

The Monte-Carlo results in Fig.~\ref{fig:epsn4} show that the ratio 
$\langle\varepsilon_3^4\rangle/\langle\varepsilon_3^2\rangle^2$ is 
slightly smaller than 2 for both Monte-Carlo KLN and Glauber models (a
fact also implied by experimental results~\cite{Bhalerao:2011ry}),
which cannot be explained by our leading-order result Eq.~(\ref{eps34}). 
The next-to-leading correction (restricted to central collisions) is derived in the Appendix. 
It is negative and scales like $1/N$. As shown in
Fig.~\ref{fig:epsn4}, it improves agreement. 

Finally, $\langle\varepsilon_1^4\rangle$ can be computed in the same
way. The leading-order result is: 
\begin{equation}
\label{eps14}
\frac{\langle\varepsilon_1^4\rangle}{\langle\varepsilon_1^2\rangle^2}
=2+\left(\frac{\varepsilon\{1,1\}}
  {\varepsilon_1\{2\}^2}\right)^2, 
\end{equation}
where 
\begin{eqnarray}
\label{eps11}
\varepsilon\{1,1\}&=&\langle\varepsilon_1^2 \cos 2\Phi_1\rangle\cr
&=& \frac{1}{N\langle r^3\rangle^2}\left(\langle r^6\cos
  2\phi\rangle
-8\varepsilon_s\langle r^2\rangle^3
+2 \varepsilon_s\langle r^2\rangle \langle r^4\rangle
\right. \cr &&\left. 
-4\langle r^2\rangle\langle r^4\cos 2\phi\rangle
- \varepsilon_s^3 \langle r^2\rangle^3\right) .
\end{eqnarray}
This quantity is negative, 
which means that the dipole asymmetry develops mostly out 
of the reaction plane for non-central collisions. 
The last term in Eq.~(\ref{eps14}) gives a positive contribution to 
the ratio
$\langle\varepsilon_1^4\rangle/\langle\varepsilon_1^2\rangle^2$,
which increases up to 3 for peripheral collisions. 
Note that $\varepsilon_1\{4\}^4$ defined by Eq.~(\ref{4cumulant}) is
negative, so that $\varepsilon_1\{4\}$ is undefined. 

The leading order result defined by Eqs.~(\ref{eps14}),
(\ref{rmsepsilon}) and (\ref{eps11}) is independent of $N$. 
It is in good agreement with Monte-Carlo results. 

Note that within our independent-source model, the ratios 
plotted in Fig.~\ref{fig:epsn4} are strictly insensitive to the
profile of a single source $\rho(r)$.

\subsection{Mixed correlations}
\label{s:correlations}

We now study the correlations between $\Phi_1$, $\Phi_2$ and
$\Phi_3$. The lowest order non-trivial correlations are 
\begin{eqnarray}
\label{correlations}
\varepsilon_{23}&\equiv& 
\varepsilon\{3,3,-2,-2,-2\}
=\left\langle\varepsilon_3^2\varepsilon_2^3\cos(6(\Phi_3-\Phi_2))\right\rangle\cr
\varepsilon_{12}&\equiv& 
\varepsilon\{1,1,-2\}
=\left\langle\varepsilon_1^2\varepsilon_2\cos(2(\Phi_1-\Phi_2))\right\rangle\cr
\varepsilon_{123}&\equiv& 
\varepsilon\{3,-2,-1\}
=\left\langle\varepsilon_3\varepsilon_2\varepsilon_1\cos(3\Phi_3-2\Phi_2-\Phi_1)\right\rangle\cr
\varepsilon_{1223}&\equiv& 
\varepsilon\{3,1,-2,-2\}
=\left\langle\varepsilon_1\varepsilon_3\varepsilon_2^2\cos(3\Phi_3+\Phi_1-4\Phi_2)\right\rangle\cr
\varepsilon_{13}&\equiv& 
\varepsilon\{3,-1,-1,-1\}
=\left\langle\varepsilon_3\varepsilon_1^3\cos(3(\Phi_3-\Phi_1))\right\rangle .
\end{eqnarray}
All these correlations were studied numerically in
Ref.~\cite{Bhalerao:2011yg}, with the exception of
$\varepsilon_{1223}$ which is new. 
We derive an analytic prediction for these quantities using 
our independent-source model. 
To leading order in $1/N$, one can replace each factor of
$\varepsilon_2 e^{\pm 2i\Phi_2}$ in Eq.~(\ref{moments}) by the
standard eccentricity $\varepsilon_s$.
The moments in Eq.~(\ref{correlations}) can then be expressed in terms
of two-point moments using Wick's theorem:
\begin{eqnarray}
\label{wick}
\varepsilon\{3,3,-2,-2,-2\}
&=&\varepsilon_s^3\varepsilon\{3,3\}\cr
\varepsilon\{1,1,-2\}
&=&
\varepsilon_s\varepsilon\{1,1\}\cr
\varepsilon\{3,-2,-1\}
&=&
\varepsilon_s\varepsilon\{3,-1\}\cr
\varepsilon\{3,1,-2,-2\}
&=&
\varepsilon_s^2\varepsilon\{3,1\}\cr
\varepsilon\{3,-1,-1,-1\}
&=&
3\varepsilon\{3,-1\}\varepsilon\{1,1\} .
\end{eqnarray}
To leading order in $1/N$, $\varepsilon\{3,3\}$ and $\varepsilon\{1,1\}$ are
given by Eqs.~(\ref{eps33}) and (\ref{eps11}). The remaining
two-point functions 
$\varepsilon\{3,-1\}$ and $\varepsilon\{3,1\}$ can be computed 
similarly:
\begin{eqnarray}
\label{eps13}
\varepsilon\{3,-1\}&=&\langle\varepsilon_3\varepsilon_1
\cos(3\Phi_3-\Phi_1)\rangle\cr 
&=& \frac{1}{N\langle r^3\rangle^2}\left(
\langle r^6\cos 2\phi\rangle
+3\varepsilon_s\langle  r^2\rangle \langle r^4\rangle
\right. \cr &&\left. 
-6\varepsilon_s\langle r^2\rangle^3
- 3\varepsilon_s^3 \langle r^2\rangle^3
-2\langle r^2\rangle\langle r^4\cos 2\phi\rangle
\right. \cr &&\left. 
+\varepsilon_s\langle r^2\rangle\langle r^4\cos 4\phi\rangle
\right)\cr
\varepsilon\{3,1\}&=&\langle\varepsilon_3\varepsilon_1
\cos(3\Phi_3+\Phi_1)\rangle\cr 
&=& \frac{1}{N\langle r^3\rangle^2}\left(
\langle r^6\cos 4\phi\rangle
+4\varepsilon_s\langle  r^2\rangle \langle r^4\cos 2\phi\rangle
\right. \cr &&\left. 
-2\langle r^2\rangle\langle r^4\cos 4\phi\rangle
+9\varepsilon_s^2 \langle r^2\rangle^3
\right) .
\end{eqnarray}
These equations define our analytic results for the moments. 

We now compare our analytic results with Monte-Carlo results. 
We first scale $\varepsilon\{n_1,\cdots,n_k\}$ by 
$\varepsilon_{n_1}\{2\}\cdots\varepsilon_{n_k}\{2\}$ to single out the
angular correlation as in Ref.~\cite{Bhalerao:2011yg}. 
We compute these ratios both with the Monte-Carlo and 
with our analytic formulas. 
To leading order in $1/N$, we use the approximation
$\varepsilon_2\{2\}\simeq \varepsilon_s$. One easily shows, using
Eq.~(\ref{wick}),  that the resulting leading-order prediction for the
ratios is independent of $N$.

\begin{figure}[ht]
\includegraphics[width=\linewidth]{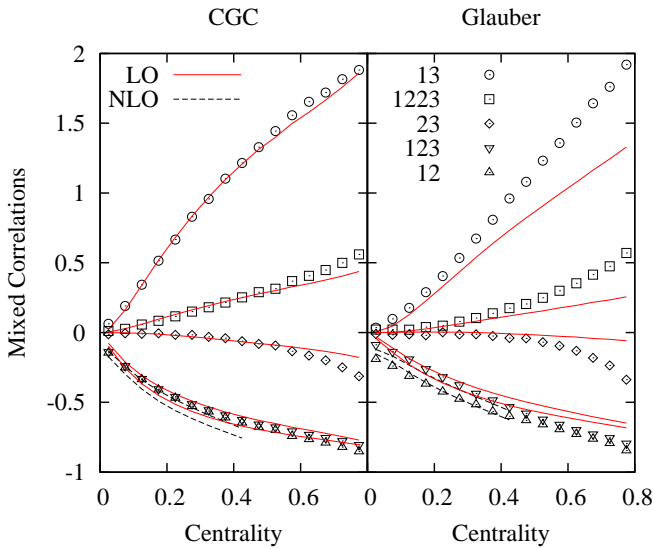}
\caption{(Color online) 
Mixed correlations versus centrality. 
From top to bottom:
$\varepsilon_{13}/(\varepsilon_1\{2\}^3\varepsilon_3\{2\})$ (labeled 13), 
$\varepsilon_{1223}/(\varepsilon_1\{2\}\varepsilon_2\{2\}^2\varepsilon_3\{2\})$
(labeled 1223),
$\varepsilon_{23}/(\varepsilon_2\{2\}^3\varepsilon_3\{2\}^2)$ (labeled 23), 
$\varepsilon_{123}/(\varepsilon_1\{2\}\varepsilon_2\{2\}\varepsilon_3\{2\})$
(labeled 123) and  
$\varepsilon_{12}/(\varepsilon_1\{2\}^2\varepsilon_2\{2\})$ (labeled 12). 
Symbols are Monte-Carlo results. 
Full lines are analytic results to leading-order in $1/N$. 
Dashed lines for $\varepsilon_{12}$ and $\varepsilon_{123}$ include
next-to-leading order corrections derived in Appendix. 
As in Fig.~\ref{fig:epsn4}, we display NLO results up to 40\%
centrality, although they are only valid for central collisions. 
}
\label{fig:correlations}
\end{figure}

Fig.~\ref{fig:correlations} displays a comparison between Monte-Carlo
results and analytic results. 
Our leading-order results explain the sign, magnitude, and 
centrality dependence of all ratios. 

The scaled $\varepsilon_{23}$ is 
much smaller than unity, in 
agreement with earlier observations that triangularity and
eccentricity are uncorrelated~\cite{Petersen:2010cw} except for peripheral 
collisions~\cite{Nagle:2010zk}. 
One thus expects that triangular flow and elliptic flow are
essentially uncorrelated, as confirmed by a recent experimental
analysis~\cite{Aamodt:2011vk}. 
Note that while data agree qualitatively with expectations, they do
not agree quantitatively with existing Monte-Carlo
models~\cite{Bhalerao:2011ry}. 

We now discuss the remaining correlations, which involve $\Phi_1$ and
are not yet measured. 
The scaled $\varepsilon_{13}$ is large for peripheral collisions,
which means that there is a strong positive correlation
between~$3\Phi_1$ and $3\Phi_3$~\cite{Staig:2010pn}. 
On the other hand, $\varepsilon_{12}$ is negative, 
which means that $\Phi_1$ is more likely to be 
perpendicular to the participant plane. 
Because of the strong correlation between $3\Phi_1$ and $3\Phi_3$,
this also explains why the mixed correlation $\varepsilon_{123}$ is 
also negative~\cite{Staig:2010pn,Teaney:2010vd}.

A closer look at Fig.~\ref{fig:correlations} reveals that 
our leading-order results for $\varepsilon_{12}$ and
$\varepsilon_{123}$ do not agree well with Monte-Carlo for central 
collisions, where both ratios are small but not zero, while our
leading-order results go smoothly to zero. 
Agreement is improved by including the next-to-leading 
term, which is of order $1/N^2$, negative, and does not vanish for
central collisions. This term is calculated in the Appendix for central
collisions.

\section{Conclusions}

We have shown that the magnitude of initial-state anisotropies
$\varepsilon_n$, their fluctuations and mutual correlations can be
understood within a simple model where fluctuations stem from
identical, independent sources, which can be viewed as ``hot spots''
scattered across the interaction region~\cite{Gyulassy:1996br,Ma:2010dv}.
The independent-source model reproduces results from both CGC-inspired and
and Glauber Monte-Carlo models. 

In our model, all information about the initial state is encoded in a
few parameters: the number of sources $N$, the profile of a single
source, and the distribution of sources across the transverse plane. 
We have considered pointlike sources for simplicity. 
The magnitudes of $\varepsilon_3$ and
$\varepsilon_1$ show that $N$ is roughly half the number of
participant nucleons, which means that participant nucleons are
correlated pairwise.  
The angular correlations between $\Phi_1$, $\Phi_2$ and $\Phi_3$  are
independent of $N$ to leading order in $1/N$. They only depend on the
distribution of sources, and are 
mostly driven by the almond shape of the overlap area between
the two nuclei, which is responsible for the large elliptic 
flow~\cite{Ollitrault:1992bk}. 

There is an ambiguity in the definition of the ``initial state''. What
is computed in Monte-Carlo models (KLN or Glauber) is typically the
distribution of energy right after the nuclei have passed through each
other. On the other hand, the relevant anisotropies for collective 
flow are the anisotropies at a somewhat later time, when the system 
reaches local equilibrium. 
The value of $\varepsilon_n$ decreases during this thermalization 
time~\cite{Qin:2010pf}. 
We have shown that this decrease is solely due to the increase
in the system size, due to the smearing of each source. 
$\varepsilon_3$ depends more strongly on the system size than
$\varepsilon_2$, which explains why it decreases more strongly during
the thermalization phase. (This 
effect was interpreted by the authors of Ref.~\cite{Qin:2010pf} as an
interference between $\varepsilon_2$ and $\varepsilon_3$.)

Within our independent-source model, ratios such as
$\langle\varepsilon_n^4\rangle/\langle\varepsilon_n^2\rangle^2$, as 
well as the various angular correlations between event planes considered 
in~\cite{Bhalerao:2011yg}, 
are independent of the number of sources to leading order in $1/N$
(with the
exception of
$\langle\varepsilon_2^4\rangle/\langle\varepsilon_2^2\rangle^2$). 
Remarkably, these ratios are also unchanged through the early-time
dynamics, and are therefore well-defined observables for
initial-state fluctuations.

\begin{acknowledgments}
We would like to thank Cl\'{e}ment Gombeaud for help with the Monte-Carlo models, and in particular for providing the code for centrality determination. 
This work is funded by ``Agence Nationale de la Recherche'' under grant
ANR-08-BLAN-0093-01 and by CEFIPRA under project 4404-2.
RSB acknowledges the hospitality of IPhT Saclay where part of this work was done.
\end{acknowledgments}

\appendix
\section{Higher-order calculations}
\label{s:NLO}

In this Appendix, we present next-to-leading order calculations in
$1/N$ for a few moments. 
Since next-to-leading order calculations are considerably more
involved than leading-order calculations, we assume azimuthal symmetry
for sake of simplicity. 
We compute three quantities for which our leading-order results vanish
in the limit of azimuthal symmetry, namely, $\varepsilon_3\{4\}$, 
$\varepsilon_{12}$ and $\varepsilon_{123}$.
 
We first compute $\varepsilon_3\{4\}$. 
Taking the recentering correction into account, the expression of
$\varepsilon_3$ is 
\begin{equation}
\varepsilon^2_3 =\frac{\{(z-\delta_z)^3 \} \{(\bar z - \delta_{\bar z})^3\}}{
\{|z - \delta_z|^3\}^2}.
\end{equation}
Treating $\delta$ as 
the expansion parameter, we can expand this expression  
to any desired order in $\delta$. 
For instance, 
$\{(z-\delta_z)^3
\}=\delta_{z^3}-3\langle
z^2\rangle\delta_z-3\delta_{z^2}\delta_z+2(\delta_z)^3$. 
We organize the calculation as follows.  Let
\begin{equation}
\varepsilon^2_3 = B + C + D + \cdots ,
\end{equation}
where $B \sim {\mathcal O} (\delta^2)$, $C \sim {\mathcal O} (\delta^3)$, etc. 
(There are no terms 
independent of $\delta$ or of order $\delta$.)  Hence
\begin{equation}
\varepsilon_3 \{2\}^2 \equiv \langle \varepsilon^2_3\rangle = \langle B \rangle 
+ \langle C \rangle + \langle D\rangle + \cdots ,
\end{equation}
and 
\begin{eqnarray}
\varepsilon_3 \{4\}^4 &\equiv& 2 \langle \varepsilon_3^2 \rangle^2 - \langle 
\varepsilon_3^4\rangle\cr
&=& 2 \langle B\rangle^2 - \langle B^2\rangle \cr
&&
+ 2\langle C\rangle^2 - 
\langle C^2\rangle \cr
&&
+ 2\langle D\rangle^2 - \langle D^2\rangle 
\cr
&& 
+ 4\langle B\rangle \langle C\rangle
 - 2\langle BC\rangle \cr
&&
+ 4\langle B\rangle \langle D\rangle - 2\langle BD\rangle \cr
&&
+ 4\langle C\rangle \langle D\rangle - 2\langle CD\rangle + \cdots. 
\end{eqnarray}
To proceed further, we make use of the expressions for $\langle \delta_f 
\delta_g\rangle$, $\langle 
\delta_f \delta_g \delta_h\rangle$, $\langle \delta_f 
\delta_g \delta_h \delta_u\rangle$, $\langle \delta_f \delta_g \delta_h 
\delta_u \delta_v\rangle$ and $\langle \delta_f \delta_g \delta_h 
\delta_u \delta_v \delta_w\rangle$ given in~\cite{Alver:2008zza}. 
(These  
expressions ignore correlations between participant positions.)  Dominant 
terms in these five quantities are of ${\mathcal O}(1/N)$,  
${\mathcal O}(1/N^2)$, ${\mathcal O}(1/N^2)$,
${\mathcal O}(1/N^3)$ and ${\mathcal O}(1/N^3)$, respectively, 
where $N$ is the number of 
participants.  
For central collisions, azimuthal symmetry allows to simplify the
previous expression:
\begin{eqnarray}
\varepsilon_3 \{4\}^4 &=& 2\langle B\rangle^2 - \langle B^2\rangle - 
\langle C^2\rangle+ 4\langle B\rangle \langle C\rangle  \cr
&&
- 
2\langle B C\rangle + 4\langle B\rangle \langle D\rangle - 
2 \langle B D\rangle.
\end{eqnarray}
After some algebra, we get
\begin{equation}
\varepsilon_3 \{4\}^4 = 
\frac{1}{N^3} 
\left[\frac{2\langle r^6\rangle^2 - \langle r^{12} \rangle
}{\langle r^3\rangle^4} + \frac{8\langle r^6 
\rangle \langle r^9\rangle}{\langle r^3\rangle^5}
- \frac{8\langle r^6\rangle^3 }{
\langle r^3\rangle^6} \right]. 
\end{equation}
A similar result has been derived previously for
$\varepsilon_2\{4\}^4$, which is also of order $1/N^3$ for independent
sources and central collisions~\cite{Alver:2008zza}. 
This scaling in $1/N^3$ is natural for the four-particle
cumulant~\cite{Borghini:2001vi}. 

Other moments can be computed in a similar way. Our results for
$\varepsilon_{123}$ and $\varepsilon_{12}$ are given 
in Appendix~\ref{s:LO}. 

\section{Summary of results}
\label{s:LO}

The expressions of $\varepsilon_n\{2\}^2$ to order $1/N$ are
given by Eqs.~(\ref{rmsepsilon}) and (\ref{eps2}). 
The higher-order cumulant $\varepsilon_1\{4\}^4$ has been
derived to order $1/N^2$:
\begin{eqnarray}
\varepsilon_1\{4\}^4&\equiv&2\langle\varepsilon_1^2\rangle^2-\langle\varepsilon_1^4\rangle\cr 
&=&
-\frac{1}{N^2\langle r^3\rangle^4}\left[\langle r^6\cos
  2\phi\rangle
-8\varepsilon_s\langle r^2\rangle^3
+2 \varepsilon_s\langle r^2\rangle \langle r^4\rangle
\right. \cr &&\left.\;\;\;\;\;\;\;\;\;\;
-4\langle r^2\rangle\langle r^4\cos 2\phi\rangle
- \varepsilon_s^3 \langle r^2\rangle^3\right]^2.
\end{eqnarray}
The expression of $\varepsilon_2\{4\}^4$ was derived
in~\cite{Alver:2008zza} up to order $1/N^3$ for central collisions:
\begin{eqnarray}
\varepsilon_2\{4\}^4&\equiv&2\langle\varepsilon_2^2\rangle^2-\langle\varepsilon_2^4\rangle\cr 
&=&\es^4 + \frac{1}{N\rsq^2}[2\es^4\av{r^4}-2\es^2\av{r^4\cos4\phi}] \nonumber \\ 
& & +\, \frac{1}{N^2} \left[ 8\es^2\frac{\av{r^6}}{\rsq^3} 
    + 4\es\frac{\av{r^6\cos2\phi}}{\rsq^3} \right. \nonumber \\
& & \left. \;\;\;\;\;\;\;\;\;\; -\, 16\es^2\frac{\av{r^4}^2}{\rsq^4}
    -16\es\frac{\av{r^4}\av{r^4\cos2\phi}}{\rsq^4} \right]  \nonumber \\
& & +\, \frac{1}{N^3} \left[ \frac{2\av{r^4}^2}{\rsq^4} -\frac{\av{r^8}}{\rsq^4}
    +\frac{8\av{r^4}\av{r^6}}{\rsq^5} \right. \nonumber \\
& & \left. \;\;\;\;\;\;\;\;\;\; -\, \frac{8\av{r^4}^3}{\rsq^6} \right] 
\label{eq:ecc4}
\end{eqnarray}
Our result for $\varepsilon_3\{4\}^4$ to order $1/N^3$ is 
\begin{eqnarray}
\varepsilon_3\{4\}^4&\equiv&2\langle\varepsilon_3^2\rangle^2-\langle\varepsilon_3^4\rangle\cr 
&=&
-\frac{1}{N^2\langle r^3\rangle^4}\left[
\langle r^6\cos 6\phi\rangle
+6\varepsilon_s \langle r^2\rangle \langle r^4 \cos 4\phi\rangle 
\right. \cr &&\left. \;\;\;\;\;\;\;\;\;\;
- 9 \varepsilon_s^3 \langle r^2\rangle^3\right]^2\cr
&&+\frac{1}{N^3\langle r^3\rangle^4} 
\left[2\langle r^6\rangle^2 - \langle r^{12} \rangle
+ \frac{8\langle r^6 
\rangle \langle r^9\rangle}{\langle r^3\rangle}\right.\cr
&&\left.\;\;\;\;\;\;\;\;\;\;- \frac{8\langle r^6\rangle^3 }{
\langle r^3\rangle^2} \right]. 
\end{eqnarray}
The 3-point mixed correlations are, to order $1/N^2$, 
\begin{eqnarray}
\varepsilon_{12}&=&
 \frac{\varepsilon_s}{N\langle r^3\rangle^2}\left[\langle r^6\cos
  2\phi\rangle
-8\varepsilon_s\langle r^2\rangle^3
+2 \varepsilon_s\langle r^2\rangle \langle r^4\rangle
\right. \cr &&\left.\;\;\;\;\;\;\;\;\;\;
-4\langle r^2\rangle\langle r^4\cos 2\phi\rangle
- \varepsilon_s^3 \langle r^2\rangle^3\right] \cr
&&+
\frac{1}{ N^2} \left[
\frac{4 \langle r^4 \rangle^2 -\langle r^8 \rangle}{ \langle r^2 
\rangle\langle r^3 \rangle^2 }
\right. \cr &&\left.\;\;\;\;\;\;\;\;\;\;
+\frac{8 \langle r^2 \rangle^3+4\langle r^6 \rangle 
-16\langle r^2 \rangle \langle r^4 \rangle
}{\langle r^3 \rangle^2} 
\right],
\end{eqnarray}
and 
\begin{eqnarray}
\varepsilon_{123}&=&
 \frac{\varepsilon_s}{N\langle r^3\rangle^2}\left[
\langle r^6\cos 2\phi\rangle
+3\varepsilon_s\langle  r^2\rangle \langle r^4\rangle
-6\varepsilon_s\langle r^2\rangle^3
\right. \cr &&\left. \;\;\;\;\;\;\;\;\;\;
- 3\varepsilon_s^3 \langle r^2\rangle^3
-2\langle r^2\rangle\langle r^4\cos 2\phi\rangle
\right. \cr &&\left. \;\;\;\;\;\;\;\;\;\;
+\varepsilon_s\langle r^2\rangle\langle r^4\cos 4\phi\rangle
\right]\cr
&&+\frac{1 }{ N^2} 
\left[\frac{2\langle r^6\rangle - 6\langle r^2\rangle \langle r^4\rangle}{ 
\langle r^3\rangle^2} 
+ \frac{3\langle r^4\rangle^2 - \langle r^8\rangle}{ 
\langle r^2\rangle \langle r^3\rangle^2}\right]. 
\end{eqnarray}
For higher-order mixed correlations, we have only carried out
leading-order calculations. 
The 4-point correlations considered in this paper are 
\begin{eqnarray}
\varepsilon_{1223}&=&
\frac{\varepsilon_s^2}{N\langle r^3\rangle^2}
\left[
\langle r^6\cos 4\phi\rangle
+4\varepsilon_s\langle  r^2\rangle \langle r^4\cos 2\phi\rangle
\right. \cr &&\left.  \;\;\;\;\;\;\;\;\;\;
-2\langle r^2\rangle\langle r^4\cos 4\phi\rangle
+9\varepsilon_s^2 \langle r^2\rangle^3
\right] 
\end{eqnarray}
and
\begin{eqnarray}
\varepsilon_{13}&=&
 \frac{3}{N^2\langle r^3\rangle^4}
\left[\langle r^6\cos
  2\phi\rangle
-8\varepsilon_s\langle r^2\rangle^3
+2 \varepsilon_s\langle r^2\rangle \langle r^4\rangle
\right. \cr &&\left.\;\;\;\;\;\;\;\;\;\;
-4\langle r^2\rangle\langle r^4\cos 2\phi\rangle
- \varepsilon_s^3 \langle r^2\rangle^3\right]\cr
&&\times
\left[
\langle r^6\cos 2\phi\rangle
+3\varepsilon_s\langle  r^2\rangle \langle r^4\rangle
-6\varepsilon_s\langle r^2\rangle^3
\right. \cr &&\left. \;\;\;\;\;\;\;\;\;\;
- 3\varepsilon_s^3 \langle r^2\rangle^3
-2\langle r^2\rangle\langle r^4\cos 2\phi\rangle
\right. \cr &&\left. \;\;\;\;\;\;\;\;\;\;
+\varepsilon_s\langle r^2\rangle\langle r^4\cos 4\phi\rangle
\right].
\end{eqnarray}
Finally, we have computed the 5-point correlation
\begin{eqnarray}
\varepsilon_{23}&=&
\frac{\varepsilon_s^3}{N\langle r^3\rangle^2}\left[
\langle r^6\cos 6\phi\rangle
+6\varepsilon_s \langle r^2\rangle \langle r^4 \cos 4\phi\rangle 
\right. \cr &&\left. \;\;\;\;\;\;\;\;\;\;
- 9 \varepsilon_s^3 \langle r^2\rangle^3\right].
\end{eqnarray}
For sake of completeness, we list all other 4- and 5-point 
correlations which can be constructed out of the first three
harmonics, and their expressions to leading order, obtained using
Wick's theorem:
\begin{eqnarray}
\varepsilon\{2,-2,3,-3\}&=&\varepsilon_2\{2\}^2\varepsilon_3\{2\}^2\cr
\varepsilon\{1,-1,2,-2\}&=&\varepsilon_1\{2\}^2\varepsilon_2\{2\}^2\cr
\varepsilon\{1,-1,3,-3\}&=&\varepsilon_1\{2\}^2\varepsilon_3\{2\}^2+\varepsilon\{3,1\}^2+\varepsilon\{3,-1\}^2\cr
\varepsilon\{1,2,3,-3,-3\}&=&\varepsilon_s(2\varepsilon_3\{2\}^2\varepsilon\{3,-1\}+\varepsilon\{3,1\}\varepsilon\{3,3\})\cr
\varepsilon\{1,1,-1,2,-3\}&=&\varepsilon_s(2\varepsilon_1\{2\}^2\varepsilon\{3,-1\}+\varepsilon\{1,1\}\varepsilon\{3,1\})\cr
\varepsilon\{1,1,-2,3,-3\}&=&\varepsilon_s(2\varepsilon\{3,1\}\varepsilon\{3,-1\}+\varepsilon_3\{2\}^2\varepsilon\{1,1\})\cr
\varepsilon\{1,2,2,-2,-3\}&=&\varepsilon_2\{2\}^2\varepsilon_{123}\cr
\varepsilon\{1,1,2,-2,-2\}&=&\varepsilon_2\{2\}^2\varepsilon_{12}\cr
\varepsilon\{1,1,1,-1,-2\}&=&3\varepsilon_1\{2\}^2\varepsilon_{12}.
\end{eqnarray}
The first of these equations means that there are no correlations
between the magnitudes of $\varepsilon_2$ and $\varepsilon_3$ to
leading order. This 
could be tested experimentally by measuring the correlation between
the magnitudes of $v_2$ and $v_3$.


\begin{thebibliography}{99}
\bibitem{Aguiar:2001ac}
  C.~E.~Aguiar, Y.~Hama, T.~Kodama and T.~Osada,
  Nucl.\ Phys.\  A {\bf 698}, 639 (2002)
  [arXiv:hep-ph/0106266].

\bibitem{Mrowczynski:2002bw}
  S.~Mrowczynski and E.~V.~Shuryak,
  Acta Phys.\ Polon.\  B {\bf 34}, 4241 (2003)
  [arXiv:nucl-th/0208052].

\bibitem{Miller:2003kd}
  M.~Miller and R.~Snellings,
  arXiv:nucl-ex/0312008.

\bibitem{Alver:2006wh}
  B.~Alver {\it et al.}  [PHOBOS Collaboration],
  Phys.\ Rev.\ Lett.\  {\bf 98}, 242302 (2007)
  [arXiv:nucl-ex/0610037].

\bibitem{CasalderreySolana:2009uk}
  J.~Casalderrey-Solana, U.~A.~Wiedemann,
  Phys.\ Rev.\ Lett.\  {\bf 104}, 102301 (2010).
  [arXiv:0911.4400 [hep-ph]].

\bibitem{Avsar:2010rf}
  E.~Avsar, C.~Flensburg, Y.~Hatta, J.~-Y.~Ollitrault, T.~Ueda,
  Phys.\ Lett.\  {\bf B702}, 394-397 (2011).
  [arXiv:1009.5643 [hep-ph]].

\bibitem{Bozek:2010pb}
  P.~Bozek,
  Eur.\ Phys.\ J.\  {\bf C71}, 1530 (2011).
  [arXiv:1010.0405 [hep-ph]].

\bibitem{Werner:2010ss}
  K.~Werner, I.~.Karpenko, T.~Pierog,
  Phys.\ Rev.\ Lett.\  {\bf 106}, 122004 (2011).
  [arXiv:1011.0375 [hep-ph]].

\bibitem{Khachatryan:2010gv}
  V.~Khachatryan {\it et al.} [ CMS Collaboration ],
  JHEP {\bf 1009}, 091 (2010).
  [arXiv:1009.4122 [hep-ex]].



\bibitem{Luzum:2010sp}
  M.~Luzum,
  Phys.\ Lett.\  B {\bf 696}, 499 (2011)
  [arXiv:1011.5773 [nucl-th]].

\bibitem{Luzum:2011mm}
  M.~Luzum,
  [arXiv:1107.0592 [nucl-th]].

\bibitem{Alver:2010gr}
  B.~Alver and G.~Roland,
  Phys.\ Rev.\  C {\bf 81}, 054905 (2010)
  [Erratum-ibid.\  C {\bf 82}, 039903 (2010)]
  [arXiv:1003.0194 [nucl-th]].

\bibitem{Adare:2011tg}
  A.~Adare {\it et al.} [ PHENIX Collaboration ],
  [arXiv:1105.3928 [nucl-ex]].


\bibitem{Aamodt:2011vk}
   {K. Aamodt \it et al.}  [ALICE Collaboration],
  Phys.\ Rev.\ Lett.\  {\bf 107}, 032301 (2011).
  [arXiv:1105.3865 [nucl-ex]].

\bibitem{Teaney:2010vd}
  D.~Teaney and L.~Yan,
  Phys.\ Rev.\ C {\bf 83}, 064904 (2011)
  [arXiv:1010.1876 [nucl-th]].

\bibitem{Luzum:2010fb}
  M.~Luzum, J.~-Y.~Ollitrault,
  Phys.\ Rev.\ Lett.\  {\bf 106}, 102301 (2011).
  [arXiv:1011.6361 [nucl-ex]].

\bibitem{Staig:2010pn}
  P.~Staig and E.~Shuryak,
  arXiv:1008.3139 [nucl-th].

\bibitem{Bhalerao:2011yg}
  R.~S.~Bhalerao, M.~Luzum, J.~-Y.~Ollitrault,
  Phys.\ Rev.\  {\bf C84}, 034910 (2011).
  [arXiv:1104.4740 [nucl-th]].

\bibitem{Andrade:2006yh}
  R.~Andrade, F.~Grassi, Y.~Hama, T.~Kodama and O.~J.~Socolowski,
  Phys.\ Rev.\ Lett.\  {\bf 97}, 202302 (2006)
  [arXiv:nucl-th/0608067].

\bibitem{Petersen:2010cw}
  H.~Petersen, G.~Y.~Qin, S.~A.~Bass and B.~Muller,
  Phys.\ Rev.\  C {\bf 82}, 041901 (2010)
  [arXiv:1008.0625 [nucl-th]].

\bibitem{Werner:2011fd}
  K.~Werner, K.~Mikhailov, Iu.~Karpenko, T.~Pierog,
  [arXiv:1104.2405 [hep-ph]].

\bibitem{Xu:2010du}
  J.~Xu and C.~M.~Ko,
  Phys.\ Rev.\  C {\bf 83}, 021903 (2011)
  [arXiv:1011.3750 [nucl-th]].

\bibitem{Xu:2011fe}
  J.~Xu, C.~M.~Ko,
  Phys.\ Rev.\  {\bf C84}, 014903 (2011).
  [arXiv:1103.5187 [nucl-th]].

\bibitem{Broniowski:2007ft}
  W.~Broniowski, P.~Bozek and M.~Rybczynski,
  Phys.\ Rev.\  C {\bf 76}, 054905 (2007)
  [arXiv:0706.4266 [nucl-th]].

\bibitem{Schenke:2010rr}
  B.~Schenke, S.~Jeon and C.~Gale,
  Phys.\ Rev.\ Lett.\  {\bf 106}, 042301 (2011)
  [arXiv:1009.3244 [hep-ph]].

\bibitem{Bhalerao:2006tp}
  R.~S.~Bhalerao and J.~Y.~Ollitrault,
  Phys.\ Lett.\  B {\bf 641}, 260 (2006)
  [arXiv:nucl-th/0607009].

\bibitem{Holopainen:2010gz}
  H.~Holopainen, H.~Niemi and K.~J.~Eskola,
  Phys.\ Rev.\  C {\bf 83}, 034901 (2011)
  [arXiv:1007.0368 [hep-ph]].

\bibitem{Qin:2010pf}
  G.~Y.~Qin, H.~Petersen, S.~A.~Bass and B.~Muller,
  Phys.\ Rev.\  C {\bf 82}, 064903 (2010)
  [arXiv:1009.1847 [nucl-th]].

\bibitem{Gardim:2011qn}
  F.~G.~Gardim, F.~Grassi, Y.~Hama, M.~Luzum, J.~-Y.~Ollitrault,
  Phys.\ Rev.\  {\bf C83}, 064901 (2011).
  [arXiv:1103.4605 [nucl-th]].

\bibitem{Qiu:2011iv}
  Z.~Qiu, U.~W.~Heinz,
  Phys.\ Rev.\  {\bf C84}, 024911 (2011).
  [arXiv:1104.0650 [nucl-th]].

\bibitem{Borghini:2001vi}
  N.~Borghini, P.~M.~Dinh and J.~Y.~Ollitrault,
  Phys.\ Rev.\  C {\bf 64}, 054901 (2001)
  [arXiv:nucl-th/0105040].

\bibitem{mckt}
Code by A.~Dumitru, a fork of MC-KLN by Y.~Nara.  Version 1.00 obtained from \url{http://physics.baruch.cuny.edu/files/CGC/CGC_IC.html}

\bibitem{Drescher:2007ax}
  H.~J.~Drescher and Y.~Nara,
  Phys.\ Rev.\  C {\bf 76}, 041903 (2007)
  [arXiv:0707.0249 [nucl-th]].

\bibitem{Albacete:2010ad}
  J.~L.~Albacete and A.~Dumitru,
  arXiv:1011.5161 [hep-ph].

\bibitem{Miller:2007ri}
  M.~L.~Miller, K.~Reygers, S.~J.~Sanders, P.~Steinberg,
  Ann.\ Rev.\ Nucl.\ Part.\ Sci.\  {\bf 57}, 205-243 (2007).
  [nucl-ex/0701025].

\bibitem{Gombeaud:2009am}
  C.~Gombeaud, J.~-Y.~Ollitrault,
  J.\ Phys.\ G {\bf G37}, 094024 (2010).
  [arXiv:0912.3623 [nucl-th]].

\bibitem{Bozek:2010wt}
  P.~Bozek, M.~Chojnacki, W.~Florkowski, B.~Tomasik,
  Phys.\ Lett.\  {\bf B694}, 238-241 (2010).
  [arXiv:1007.2294 [nucl-th]].

\bibitem{Alver:2008aq}
  B.~Alver, M.~Baker, C.~Loizides and P.~Steinberg,
  arXiv:0805.4411 [nucl-ex].

\bibitem{Hirano:2005xf}
  T.~Hirano, U.~W.~Heinz, D.~Kharzeev, R.~Lacey, Y.~Nara,
  Phys.\ Lett.\  {\bf B636}, 299-304 (2006).
  [nucl-th/0511046].

\bibitem{Lappi:2006xc}
  T.~Lappi, R.~Venugopalan,
  Phys.\ Rev.\  {\bf C74}, 054905 (2006).
  [nucl-th/0609021].

\bibitem{Aamodt:2010pa}
  K. Aamodt {\it et al.} [ The ALICE Collaboration ],
  Phys.\ Rev.\ Lett.\  {\bf 105}, 252302 (2010).
  [arXiv:1011.3914 [nucl-ex]].


\bibitem{Voloshin:2007pc}
  S.~A.~Voloshin, A.~M.~Poskanzer, A.~Tang and G.~Wang,
  Phys.\ Lett.\  B {\bf 659}, 537 (2008)
  [arXiv:0708.0800 [nucl-th]].

\bibitem{Alver:2008zza}
  B.~Alver {\it et al.},
  Phys.\ Rev.\  C {\bf 77}, 014906 (2008)
  [arXiv:0711.3724 [nucl-ex]].

\bibitem{Bhalerao:2011ry}
  R.~S.~Bhalerao, M.~Luzum, J.~-Y.~Ollitrault,
  [arXiv:1106.4940 [nucl-ex]].

\bibitem{Nagle:2010zk}
  J.~L.~Nagle and M.~P.~McCumber,
  Phys.\ Rev.\  C {\bf 83}, 044908 (2011)
  [arXiv:1011.1853 [nucl-ex]].

\bibitem{Gyulassy:1996br}
  M.~Gyulassy, D.~H.~Rischke, B.~Zhang,
  Nucl.\ Phys.\  {\bf A613}, 397-434 (1997) 
  [nucl-th/9609030].

\bibitem{Ma:2010dv}
  G.~-L.~Ma, X.~-N.~Wang,
  Phys.\ Rev.\ Lett.\  {\bf 106}, 162301 (2011).
  [arXiv:1011.5249 [nucl-th]].

\bibitem{Ollitrault:1992bk}
  J.~-Y.~Ollitrault,
  Phys.\ Rev.\  {\bf D46}, 229-245 (1992).

\end{thebibliography}
\end{document}